\documentstyle[12pt,epsfig,worldsci]{article}
\newdimen\rotdimen
\def\vspec#1{\special{ps:#1}}
\def\rotstart#1{\vspec{gsave currentpoint currentpoint translate
   #1 neg exch neg exch translate}}
\def\rotfinish{\vspec{currentpoint grestore moveto}}
\def\rotr#1{\rotdimen=\ht#1\advance\rotdimen by\dp#1%
   \hbox to\rotdimen{\hskip\ht#1\vbox to\wd#1{\rotstart{90 rotate}%
   \box#1\vss}\hss}\rotfinish}
\def\rotl#1{\rotdimen=\ht#1\advance\rotdimen by\dp#1%
   \hbox to\rotdimen{\vbox to\wd#1{\vskip\wd#1\rotstart{270 rotate}%
   \box#1\vss}\hss}\rotfinish}%
\def\rotu#1{\rotdimen=\ht#1\advance\rotdimen by\dp#1%
   \hbox to\wd#1{\hskip\wd#1\vbox to\rotdimen{\vskip\rotdimen
   \rotstart{-1 dup scale}\box#1\vss}\hss}\rotfinish}%
\def\rotf#1{\hbox to\wd#1{\hskip\wd#1\rotstart{-1 1 scale}%
   \box#1\hss}\rotfinish}%
\font\tenrm=cmr10

\begin{document}
\renewenvironment{thebibliography}[1]
  { \begin{list}{\arabic{enumi}.}
    {\usecounter{enumi} \setlength{\parsep}{0pt}
     \setlength{\itemsep}{3pt} \settowidth{\labelwidth}{#1.}
     \sloppy
    }}{\end{list}}

\parindent=1.5pc

\begin{flushright}
OCIP/C 94-6 \\
hep-ph/9407345 \\
\end{flushright}

\begin{center}{{\bf PROBING THE HEAVY QUARK CONTENT OF THE PHOTON\\
               \vglue 3pt
		USING b TAGGING AT ELECTRON PHOTON COLLIDERS} \\
\vglue 1.0cm
{K. ANDREW PETERSON
\footnote{Talk presented by K. Andrew Peterson at
{\it MRST-94: What Next? Exploring the Future of High-Energy Physics,}
McGill University, Montr\'eal, Qu\'ebec, Canada, May 11-13, 1994.}
\footnote{Address after 1 August 1994:  Department of Physics,
Memorial University of Newfoundland, St. John's, NF, CANADA, A1B 3X7.}
, M. A. DONCHESKI and STEPHEN GODFREY}\\
\baselineskip=14pt
{\it Physics Department, Carleton University , 1125 Colonel By Dr.,}\\
\baselineskip=14pt
{\it Ottawa, Ontario, Canada.  K1S 5B6}\\
\vglue 0.8cm
{\tenrm ABSTRACT}}
\end{center}
{\rightskip=3pc
 \leftskip=3pc
 \tenrm\baselineskip=12pt
 \noindent
We study the prospects for probing the quark content of the photon
using b tagging at high energy electron-photon colliders.
We find that heavy quark tagging provides a sensitive and effective
probe of the quark content of the photon.  Using a 500 GeV $e^+e^-$\
NLC in electron-photon mode, the cross section for
$eb_{\gamma}\rightarrow eb$\ can be measured to 10\% with b tagging.
This is sufficient to differentiate between the various photon structure
functions.

\vglue 0.8cm}
%
%
{\bf\noindent 1. INTRODUCTION}
\vglue 0.2cm
\baselineskip=14pt

There is a growing interest in the hadronic content of the
photon\cite{witten}
as both a test of QCD and as a background to precision measurements
of electroweak parameters.  Although several theoretical predictions
for the hadronic content of the photon exist in the literature
\cite{do,dg,grv,lac}, the experimental data \cite{expt}
 is too sparse at this point
to significantly constrain the theory.  This paper will present a
novel method of constraining the $b$ content of the photon by using
electron photon colliders to measure hard scattering of $b$-quarks
from the photon with the electron beam.  The advantage of investigating
$b$-quarks will be that the $b$-quarks can be tagged, giving
us a method to isolate the $b$-quark content of the photon
from the light quark and gluon content.

Since the photon structure functions grow with increasing
momentum transfer, $Q^2$, it is advantageous to work with
the largest possible centre of mass energy in our collisions
of electrons and photons.  If we are using effective photon
theory, where the photons are produced through the bremstrahlung
process, the photons are soft and the effective centre of mass
energy will be small.
Another possibility is to convert a high
energy $e^+e^-$\ collider into an $e\gamma$\ collider by
Compton backscattering a high intensity laser off one of the
electron beams\cite{gin.83}.  Here the effective centre of
mass energy for electron-photon collisions is very near the
centre of mass energy of the original $e^+e^-$\ beam.
The potential of $e\gamma$\ colliders to study the hadronic
content of the photon has already been studied by other
researchers\cite{halzen,bawa}, but this was in the
context of dijet production where it is difficult to
isolate the quark content from the gluonic content.

Using a 500 GeV $e^+e^-$\ collider converted to an
$e\gamma$\ collider by backscattering laser photons,
we find that the cross-section for the process
$e + \gamma \rightarrow e + b + X$ can be measured to an
accuracy of 9\%, allowing one to discriminate between the
various photon structure functions.

\begin{figure}
\begin{center}
\mbox{\epsfig{figure=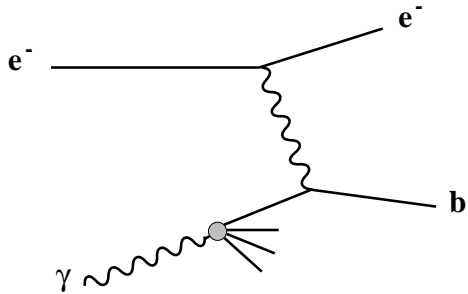,height=4cm}}
\end{center}
\caption{
The Feynman diagram for the process
$e + \gamma \rightarrow e + b + X$
}
\end{figure}
\vglue 0.6cm
{\bf\noindent 2. CALCULATIONS AND RESULTS}
\vglue 0.4cm

The process that we are interested in,
$e + \gamma \rightarrow e + b + X$,
is shown in figure 1.  It is straight forward to calculate the
cross section for the sub-process $e + b \rightarrow e + b$.  To obtain
a physical cross section, we then fold in the sub-process cross
section with the distribution functions for the hadronic content
of the photon $(b^\gamma(x,Q^2))$
and the backscattered laser $(f_{\gamma}(\tau))$:
\begin{equation}
\sigma(e + \gamma \rightarrow e + b + X)
 = \int d\tau f_{\gamma}(\tau)
   \int dx b^\gamma(x,Q^2) \sigma(e + b \rightarrow e + b)({\hat{s}}),
\end{equation}
where the effective centre of mass energies squared for
$e\gamma$ and $eb$ collisions are
\begin{equation}
\tilde{s} = \tau s
 \;\;\; {\rm and} \;\;\;
\hat{s} = x \tilde{s} = x \tau s,
\end{equation}
respectively.
Unless explicitly stated otherwise, we will take $Q^2 = {\hat{s}}$\
throughout this paper.

In the $eb$ centre of mass frame, our signal will be a electron
back to back with a $b$ jet. Since the $eb$ centre of mass frame is
boosted along the incoming electron direction,
in the lab frame this will manifest
itself as an electron and a $b$ jet with equal transverse momentum,
balanced in azimuthal angle.  To insure that both our $b$ jet and
electron will be detected, we impose beam line cuts of
$|\cos\theta_b| < 0.85$ on the angle
of the $b$ jet and
$|\cos\theta_e| < \cos 10^0$
on the angle of the electron.

The major backgrounds to this process will be:
\begin{equation}
e + \gamma \rightarrow e + b + \overline{b},
\end{equation}
\begin{equation}
e^+ + e^- \rightarrow \gamma + b + \overline{b},
\end{equation}
and
\begin{equation}
\gamma + g_\gamma \rightarrow b + \overline{b},
\end{equation}
where in each case, the $b$ is successfully tagged and all the observable
decay products of the $\overline{b}$ go down the beam pipe with the
exception of the $e^+$, which is mistakenly identified as the $e^-$.
If charge identification is possible, then all these backgrounds
will immediately disappear.  If not, then we shall demonstrate that
they are still manageable once we impose that the transverse momentum
of the electron ($p_{Te}$) must balance the transverse momentum
of the $b$ jet ($p_{Tb}$).

\begin{figure}[tb]
\begin{center}
\vspace{-5cm}\setbox1=\vbox{
\mbox{\epsfig{figure=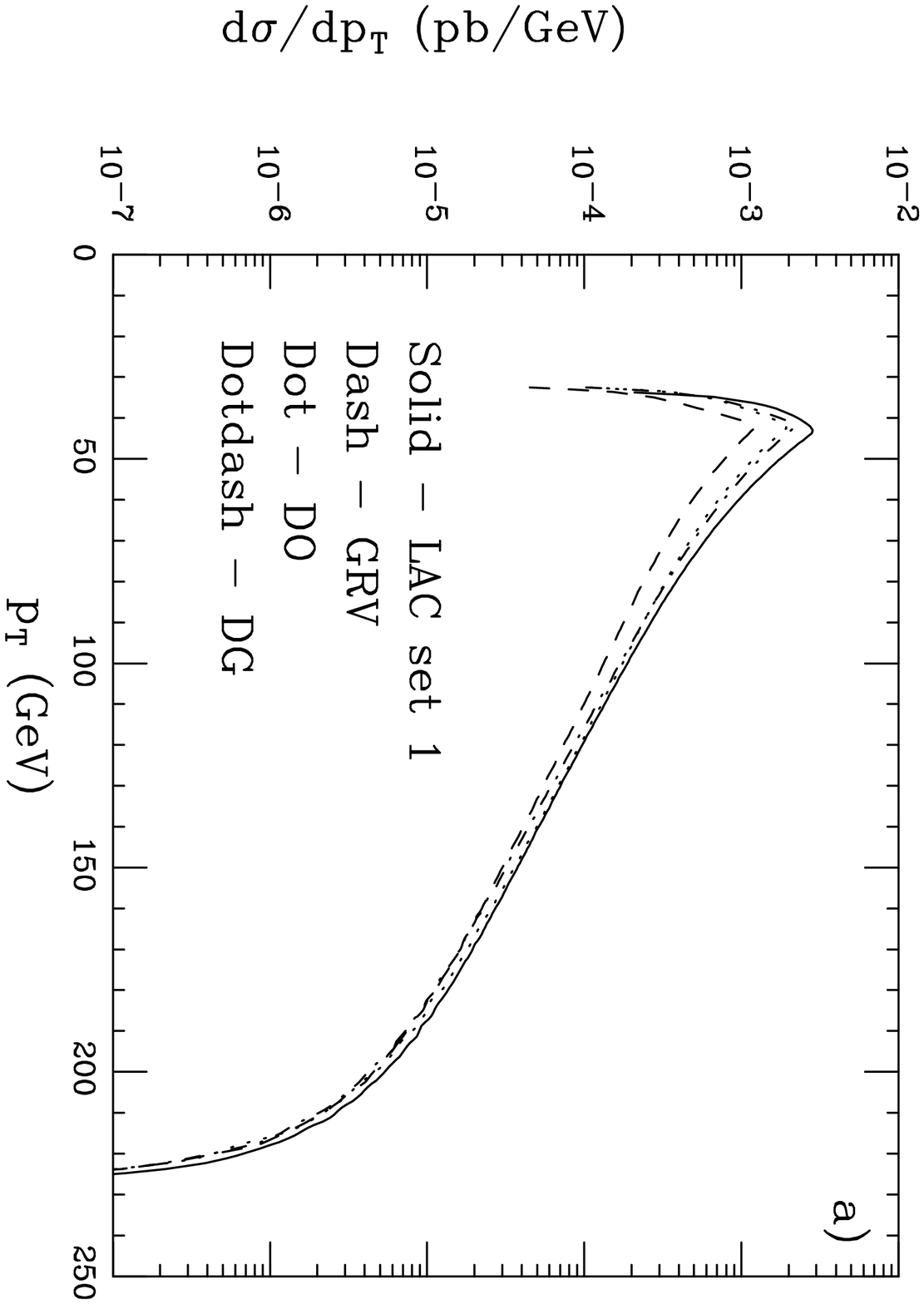,height=14cm}}
}
\rotl{1}
\vspace{-6cm}
\setbox1=\vbox{
\mbox{\epsfig{figure=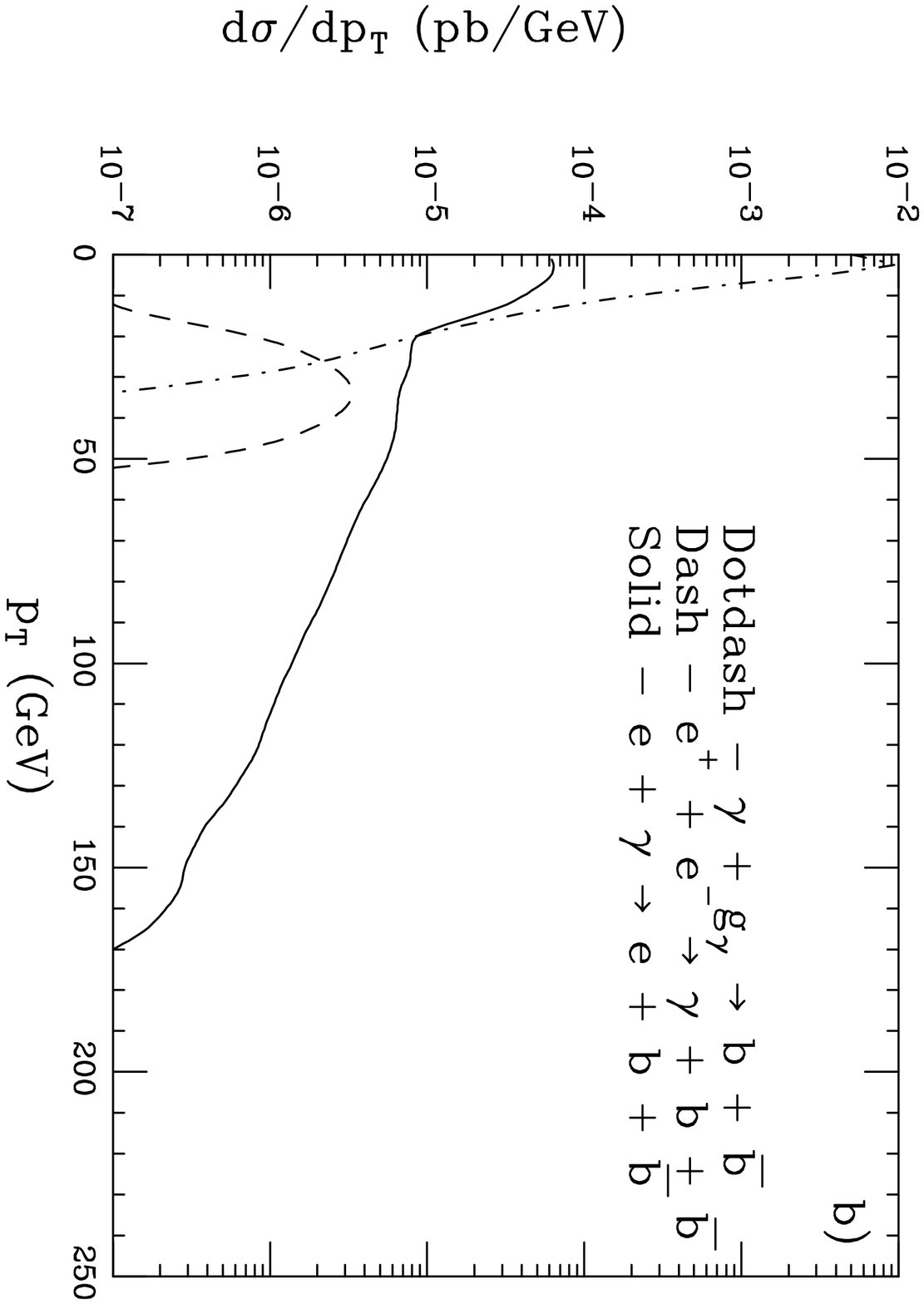,height=14cm}}
}
\rotl{1}
\vspace{-4cm}
\end{center}
\caption{
Transverse momentum distributions for the signal and backgrounds.
a) Transverse momentum distributions for different structure functions:
Solid line is LAC set 1\protect\cite{lac};
Dashed line is GRV set\protect\cite{grv};
Dotted line is DO set\protect\cite{do};
Dot-dashed line is DG set\protect\cite{dg}.
b) Transverse momentum distributions for the major backgrounds:
Solid line is $ e + \gamma \rightarrow e + b + \overline{b}$;
Dashed line is $ e^+ + e^- \rightarrow \gamma + b + \overline{b}$;
Dot-dashed line is $ \gamma  + g_{\gamma} \rightarrow b + \overline{b}$.
}
\end{figure}

In figure 2a we show $p_T$ distributions of the $b$-quark for the
signal using various different photon structure functions.
This can be compared with figure 2b, the $p_T$ distributions
of the major backgrounds where it has been imposed that
the $b$-quark and electron transverse momentums balance,
\begin{equation}
|p_{Tb} - p_{Te}| \leq 10 GeV.
\end{equation}
For the gluonic background we have used the Duke-Owen structure
functions and $Q^2 = \hat{s}(\gamma,g)$.
As one can see, the backgrounds are all small compared
to the signal, except possibly in the low $p_T$ regions.
These low $p_T$ regions can be eliminated by a simple $p_T$
cut of 25 GeV, since our signal has no cross section in that
region.

The total cross sections for the various photon structure functions
are given in table 1 together with the event rates assuming an
integrated luminosity of $50 (fb)^{-1}$\ and a $b$ reconstruction
efficiency of 10\%.  Assuming a purely statistical error in the
measurement of the cross section, we see that cross section for
the Duke Owens structure functions can be evalulated to 9\% at
the 95\% confidence level (2 sigma).  Since that is
the level of discrepancy between the various structure functions,
it should be possible to discriminate between them.

\begin{table}[h]
\begin{center}
\begin{tabular}{||l|c|c|c|c||}
\hline
\hline
Structure Function&$\sigma$ (fb)&Events&$\delta\sigma (95\% c.l.)$
&$(\sigma_i - \sigma_{D.O.})/{\sigma_{D.O.}}$\\\hline
D.O.\cite{do}&102&510&9\%&--\\
D.G.\cite{dg}&108&540&9\%&6\%\\
G.R.V.\cite{grv}&68&340&11\%&33\%\\
L.A.C.1\cite{lac}&138&690&8\%&34\%\\
\hline
\hline
\end{tabular}
\end{center}
\caption{Cross sections for the various structure functions together
with the corresponding number of events assuming
$\int {\cal L} dt = 50 fb^{-1}$ and a 10\% $b$ reconstruction
efficiency.  The third column is the accuracy to which the cross section
could be measured at the 95\% c.l. and the fourth column is the
\% difference between the various structure functions.}
\end{table}

\begin{figure}
\begin{center}
\vspace{-5cm}
\setbox1=\vbox{
\mbox{\epsfig{figure=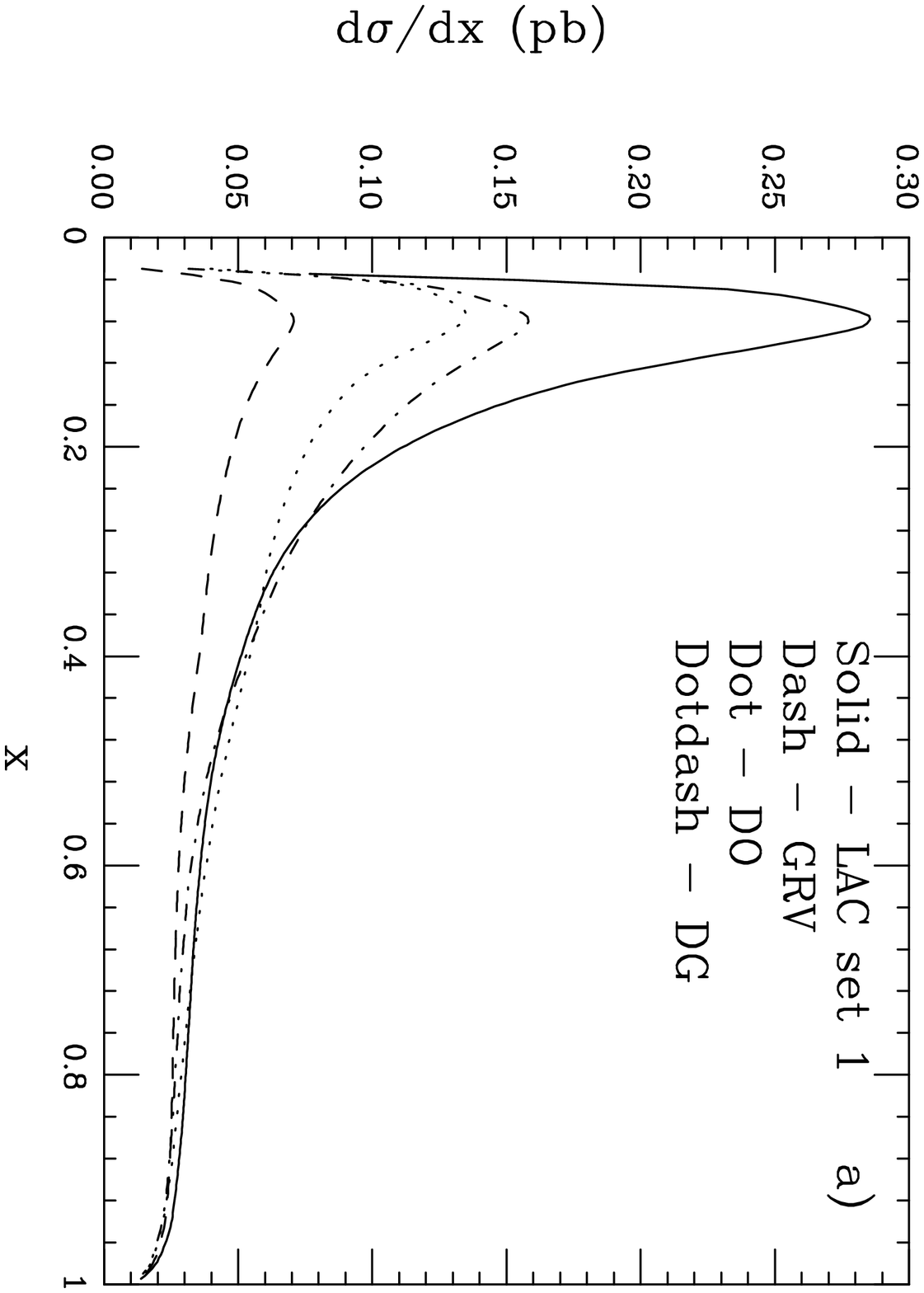,height=14cm}}
}
\rotl{1}
\vspace{-6cm}
\setbox1=\vbox{
\mbox{\epsfig{figure=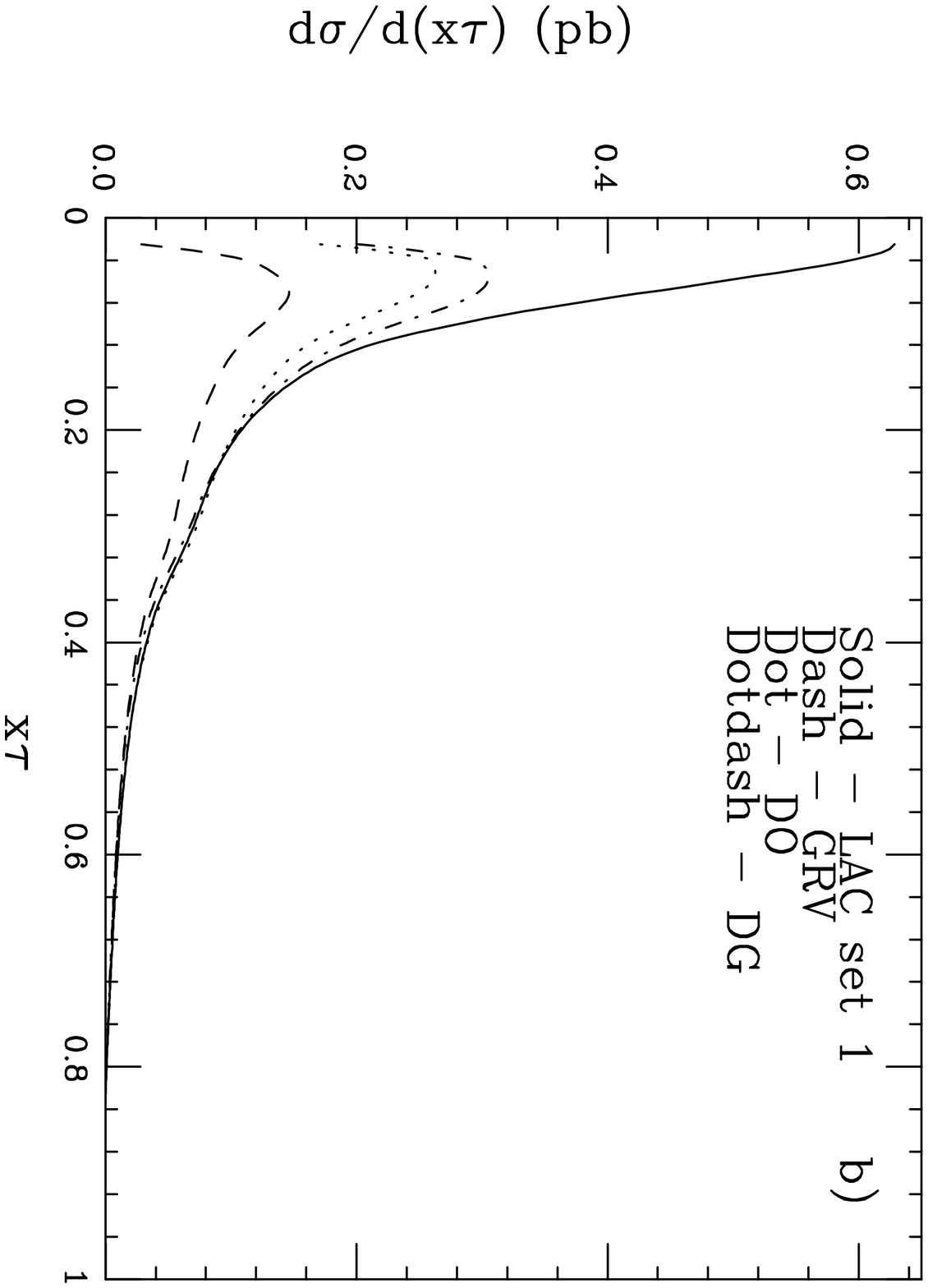,height=14cm}}
}
\rotl{1}
\vspace{-4cm}
\end{center}
\caption{
a) $x$ distributions for the various structure functions.
b) $x\tau$ distributions for the various structure functions.
Solid line is LAC set 1\protect\cite{lac};
Dashed line is GRV set\protect\cite{grv};
Dotted line is DO set\protect\cite{do};
Dot-dashed line is DG set\protect\cite{dg}.
}
\end{figure}

A more useful constraint on the structure functions could be obtained
if one can measure the $x$ distribution of the cross-section.
This distribution is shown in figure 3a.  Since $x$ is not
a directly measurable quantity, this distribution would have to be
obtained by deconvoluting the $x\tau$ distribution with the distribution
function for backscattered lasers.
This $x\tau$ distribution is given in figure 3b.
Work on the analysis of these distributions is
currently in progress \cite{us}.

\vglue 0.6cm
{\bf\noindent 2. CONCLUSIONS}
\vglue 0.4cm

We have seen that one can measure the cross section for
$e + \gamma \rightarrow e + b + X$ to 9\%.  This should allow one
to differentiate between the various structure functions
parametrizing the hadronic content of the photon.
We hope to be able to further constrain these structure functions
through the use of the $x$ distributions of their cross sections.
\vglue 0.6cm
{\bf \noindent Acknowledgements \hfil}
\vglue 0.4cm
This research was funded in part by a grant from the Natural Science
and Engineering Council of Canada.  The work of M.A.D. was funded
through a NSERC Canada International Fellowship.
\vglue 0.6cm
{\bf\noindent References \hfil}
\vglue 0.4cm

\end{document}